\documentclass[10pt,onecolumn,letterpaper]{article}

\usepackage[top=1in, bottom=1in, left=1in, right=1in]{geometry}

\usepackage{graphicx}

\begin{document}

\title{Efficient Deterministic Replay Using Complete Race Detection}

\date{}
\author{
Qi Guo, Yunji Chen, Tianshi Chen, and Ling Li\\\\
Institute of Computing Technology, Chinese Academy of Sciences\\
Loongson Technologies Corporation Limited}

\maketitle
\begin{abstract}
Data races can significantly affect the executions of multi-threaded
programs. Hence, one has to recur the results of data races to
deterministically replay a multi-threaded program. However, data
races are concealed in enormous number of memory operations in a
program. Due to the difficulty of accurately identifying data races,
previous multi-threaded deterministic record/replay schemes for
commodity multi-processor system give up to record data races
directly. Consequently, they either record all shared memory
operations, which brings remarkable slowdown to the production run,
or record the synchronization only, which introduces significant
efforts to replay.

Inspired by the advances in data race detection, we propose an
efficient software-only deterministic replay scheme for commodity
multi-processor systems, which is named RacX. The key insight of
RacX is as follows: although it is NP-hard to accurately identify
the existence of data races between a pair of memory operations, we
can find out all potential data races in a multi-threaded program,
in which the false positives can be reduced to a small amount with
our automatic false positive reduction techniques. As a result, RacX
can efficiently monitor all potential data races to
deterministically replay a multi-threaded program.

To evaluate RacX, we have carried out experiments over a number of
well-known multi-threaded programs from SPLASH-2 benchmark suite and
large-scale commercial programs. RacX can precisely recur production
runs of these programs with value determinism. Averagely, RacX
causes only about 1.21\%, 1.89\%, 2.20\%, and 8.41\% slowdown to the
original run during recording (for 2-, 4-, 8- and 16-thread
programs, respectively). The soundness, efficiency, scalability, and
portability of RacX well demonstrate its superiority.

\end{abstract}


\section{Introduction}

Due to the importance of determinism, deterministic replay has been
widely employed in many areas, including
debugging~\cite{King05ATC,Tucek07SOSP}, fault
tolerance~\cite{Bressoud96TCS}, intrusion
detection~\cite{Dunlap02OSDI}, workload
capture~\cite{Narayanasamy06SIGMETRICS}, performance
prediction~\cite{Zhai10PPoPP}, and so on. Briefly speaking, a
deterministic record/replay scheme records the non-deterministic
events in the \emph{production run} as logs, and uses the recorded
logs to force the \emph{replay run} recurring the behavior of the
production run.

For uni-processor systems, the non-deterministic events mainly
include the I/O inputs (e.g., keyboard, network, and files, etc.)
and interrupts, which are uncertain in value or timing. These
external non-deterministic events can be handled with relatively low
overheads~\cite{Dunlap02OSDI,Guo08OSDI}. As a result, deterministic
replay scheme for uni-processor systems has been a mature technique,
and it has already been accepted by industry~\cite{GDB2009}.

For multi-processor systems, there are two additional types of
non-deterministic events which may break the determinism:
synchronization and data races\footnote{The concrete definition of
data race can be found in~\cite{Adve10CACM}: There is data race
between a pair of \emph{conflicting} memory operations (we say that
two memory operations \emph{conflict} if they access the same memory
location and at least one of them is a write) in a program, if there
is a sequentially consistent execution (i.e., a program-ordered
interleaving of operations of the individual threads) of the
program, in which these two memory operations occur next to each
other.}. The common wisdom so far is that synchronization can be
handled easily for deterministic replay schemes~\cite{Ronsse99TCS},
while data races are intractable. Hence, programmers are still
crying out for a practical deterministic record/replay scheme for
commodity multi-processor systems.

Interestingly, the intractability of data race does not come from
the amount of data races. In fact, \emph{the number of data race is
quite few in real-world multi-threaded programs since most memory
operations have been well ordered by
synchronization}~\cite{Marino09PLDI}. The trouble is that data races
are concealed in enormous number of memory operations, while it is
NP-hard to identify whether there exists data race between a pair of
memory operations~\cite{Netzer90ICPP}. The difficulty in accurately
identifying data race makes researchers giving up to directly
record/replay data races. Some deterministic replay schemes would
rather to record/replay all memory operations to guarantee that each
data race behaves the same in the production run and the replay run.
So far, these schemes bring remarkable slowdown to the production
run, e.g., $36X$-$146X$ in PinPlay~\cite{Patil10CGO} and
$1.5X$-$8.7X$ in SMP-ReVirt~\cite{Dunlap08VEE}. Some other
deterministic replay schemes do not record any memory operation at
all (yet they still record all synchronization orders) to reduce the
slowdown to the production run. However, these schemes either need
tedious efforts to infer the behaviors of all memory operations,
which may consume more than $500$x of the program execution
time~\cite{Altekar09SOSP}, or repeatedly replay the program (or
program section) until the output of a replay run happen to be
identical with of the production run, while the replay times may be
exponential with respect to the number of data races in the program
(or program section)~\cite{Park09SOSP,Lee09MICRO}. Due to the
inefficiency in either record or replay, few deterministic replay
scheme for commodity multi-processor system has been widely accepted
by industry.

In this paper, we propose an efficient deterministic replay scheme
for commodity multi-processor systems, which is named RacX. The key
insight of RacX is inspired by recent advances in the investigations
of data race detection: \emph{although it is NP-hard to accurately
identify whether there is data race between a pair of memory
operations, it is not hard to find out all real data races in a
multi-threaded program with some false positives}. Thus, we can
trace the order of all the potential data races (including real
races and false positives) and synchronization to deterministically
replay multi-threaded programs.

Concretely, RacX employs a compile time data race detector
\emph{Analyzer}, which utilizes R\textsc{elay}~\cite{Voung07FSE} as
the basic infrastructure, to statically identify all the potential
data races in multi-threaded programs. The \emph{Analyzer}
incorporates a series of automatic false positive reduction
techniques proposed in this paper, which can resolve the false
positives caused by unlocked initialization and false sharing on
array. Hence, the potential races reported by \emph{Analyzer}
include all real data races and a small amount of false positives,
which involve quite few memory operations. As a result, RacX can use
static instrumentation to track the orders of all the potential data
races and synchronization to efficiently replay multi-threaded
programs with high fidelity.

We evaluate RacX over several well-known multi-threaded benchmarks,
such as SPLASH-2, Apache, desktop applications and so on. These
programs cover different application fields with various scales. All
of these programs can be faithfully replayed by RacX on a commodity
16-core Linux server. Averagely, RacX causes only about 1.44\%,
2.53\%, 3.07\%, and 4.61\% slowdown to the original run when running
in the recording mode for 2-, 4-, 8- and 16- thread programs
respectively.

In general, the superiority of RacX, which is the first
deterministic replay scheme based on complete data race detection,
can be well demonstrated by its soundness, efficiency, scalability,
and portability:
\begin{itemize}
\item \emph{Soundness}: RacX implements instrumentation to all real data races and synchronization,
which can fully determine the logical behavior of a multi-threaded
program execution. Therefore, RacX provides the value determinism
that is more faithful than the output determinism provided by many
state-of-the-art deterministic replay schemes 
\item \emph{Efficiency}: Benefited by the false positive elimination techniques,
RacX only needs to monitor a small portion of all the memory
operations, thus causes the minimal production run slowdowns among a
number of state-of-the-art deterministic replay schemes for
commodity multi-processor systems. 
Furthermore, in
RacX, it is no necessary to infer the behaviors of memory
operations, or repeatedly replay to match a desirable replay run.
\item\emph{Scalability}: RacX exhibits excellent scalability with
respect to the program size and the number of threads, as evidenced
 by that it can handle commercial multi-threaded programs with nearly millions of lines of code and 16-thread programs.
\item\emph{Portability}: Different from many deterministic replay schemes that need specific software
environments~\cite{Dunlap08VEE,Guo08OSDI,Olszewski09ASPLOS,Lee10ASPLOS,Veeraraghavan11ASPLOS}
or hardware
extensions~\cite{Hower08ISCA,Montesinos08ISCA,Voskuilen10ISCA}, RacX
has no restriction on library, operating system, and hardware.
Hence, RacX can be employed by any commodity computer system.
\end{itemize}

The rest of this paper proceeds as follows.
Section~\ref{sec:overview} introduces the overall framework of RacX.
Section~\ref{sec:race} introduces the infrastructure of the complete
data race detector employed by RacX. Section 4 and Section 5
introduce our automatic false positive reduction techniques in
details, which are crucial for the efficiency of RacX. Section 6
introduces the implementations of the deterministic record/replay.
Section~\ref{sec:experiments} gives the experimental results over a
number of representative multi-threaded benchmarks.
Section~\ref{sec:related} briefly reviews some related work.
Finally, Section~\ref{sec:conclusion} concludes this paper.

\section{Design Overview}~\label{sec:overview}
\begin{figure} [htbg]
\centering
\includegraphics[width=0.7\textwidth]{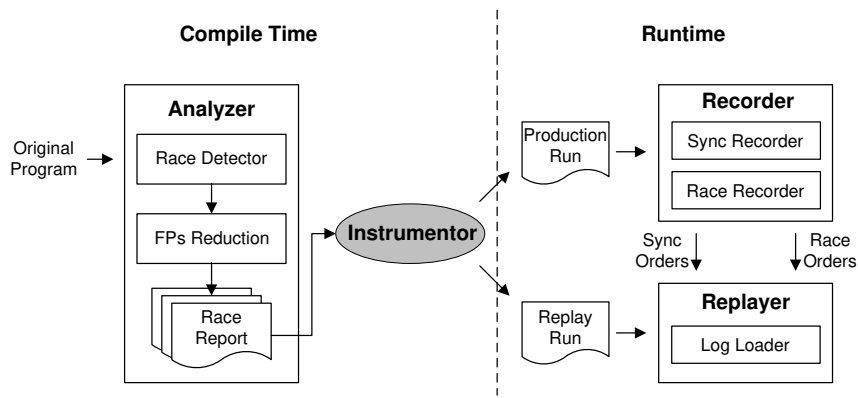}
\caption{The framework of RacX. RacX consists of four main parts as:
the \emph{Analyzer}, the \emph{Instrumentor}, the \emph{Recorder},
and the \emph{Replayer}. RacX finds all potential data races by the
\emph{Analyzer}, instruments all potential data races and
synchronization by the \emph{Instrumentor}, records their orders as
logs by the \emph{Recorder}, and replays the logs by the
\emph{Replayer}.}\label{fig:framework}
\end{figure}

In this section, we elaborate the overall framework of RacX. As
shown in Figure~\ref{fig:framework}, RacX is a compiler and runtime
environment that can execute on a commodity multi-processor system.
During the compile time, RacX employs a static data race detector
\emph{Analyzer} to identify all potential data races (including all
real data races and false positives). Due to the lack of runtime
information, state-of-the-art static data race detectors have to
make conservative assumption on the source codes, which leads
considerable false positives. Hence, we propose several false
positive elimination techniques, which include initialization
pruning and array cross-range checking, to filter
performance-critical false positives. As a result, the potential
data races in the final race report can be reduced to a quite small
amount with respect to all memory operations in a multi-threaded
program. Then, the \emph{Instrumentor} inserts order-tracking codes
to all potential data races and synchronization in the original
program.

During runtime, the \emph{Recorder} and the \emph{Replayer}
cooperate to deterministically record/replay the instrumented
program. In the production run, the \emph{Recorder} calls two
sub-recorders to log the orders of synchronization operations and
memory references involved in reported potential data races,
respectively. According to the recorded orders in the logs, the
\emph{Replayer} can control the interleaving among different threads
in the replay run with the help of the instrumented codes provided
by the \emph{Instrumentor}. By enforcing the identity of the
synchronization orders and race orders in the production run and the
record run, we can ensure to reproduce high-fidelity replica of the
production run when needed.

In general, the \emph{Analyzer} plays a key role in RacX, since it
decides the efficiency and the soundness of RacX. In comparison with
the \emph{Analyzer}, the \emph{Instrumentor}, the \emph{Recorder}
and the \emph{Replayer} are relatively straightforward. Hence, we
will introduce detail implementation of our \emph{Analyzer} with the
subsequent sections.

\section{Complete Data Race Detection}~\label{sec:race}
As we have mentioned, the soundness of RacX depends heavily on the
completeness of the \emph{Analyzer}, since even the miss of one real
data race may cause the replay run to exhibit different behavior
with the production run. Hence, it is important to select an
appropriate data race detection approach for the \emph{Analyzer}. In
general, there are two kinds of data race detection approaches,
including dynamic approaches and static approaches. The dynamic
approaches find data races in the executions of the program. They
may neglect some real data races, but they have few (even no) false
positive. The static approaches analyze the program at compile time.
Most of them do not neglect any real data races, but they may have
thousands of (even more) false positives. Since we must find out all
real data races exhaustively for deterministic record/replay, the
\emph{Analyzer} employs static data race detector as its basic
infrastructure.

On the other hand, the efficiency of RacX also depends on the
accuracy of the \emph{Analyzer}. Traditionally, static data race
detectors have a lot of false positives. Apparently, the fewer false
positives, the fewer memory operations to be instrumented, the fewer
slowdowns to the original run in record and replay mode. Hence, we
also manage to reduce the number of false positives of the
\emph{Analyzer} to improve the efficiency of RacX.

\subsection{Relay Background}
Several static data race detectors have been proposed in the last
decade~\cite{Flanagan00PLDI,Boyapati02OOSPLA,Engler03SOSP,Pratikakis06PLDI,Naik06PLDI,Voung07FSE}.
Most of them are claimed to be able to find all real data races
(with false positives) from the source code. In RacX, we consider to
adopt a context- and flow-sensitive static data race detector, which
is named R\textsc{elay}~\cite{Voung07FSE}, as the basic
infrastructure of our \emph{Analyzer}, since R\textsc{elay} can
scale to programs with millions of lines of code.

The core algorithm of many data race detectors (including
R\textsc{elay}) is based on the \emph{lockset}
algorithm~\cite{Savage97TCS}, i.e., it computes or monitors the set
of locks imposed on each shared variable, and intersects the lock
sets of different (potential) threads to determine race on each
shared variable.

In comparison with other static lockset-based detectors,
R\textsc{elay} has a better scalability with respect to the program
size. The main reason is that R\textsc{elay} adopts a
divide-and-conquer methodology, that is, it first analyzes the
information of each function independent of its calling context, and
then summarizes these information together to find potential data
races. As a consequence, it can even deal with programs with
millions of lines of code.

\begin{figure}
\centering
\includegraphics[width=0.4\textwidth]{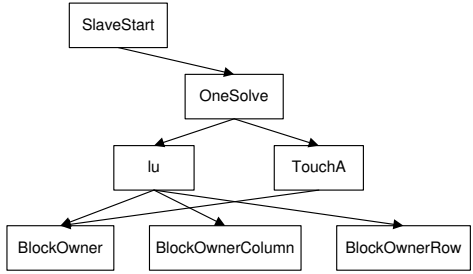}
\caption{An example of the working flow of
R\textsc{elay}.}\label{fig:relay}
\end{figure}

Technically, for each function, R\textsc{elay} computes the relative
locksets and the guarded access sets. The relative lockset describes
the intra-function changes of the lock ownerships from the function
entry. It can be formulated as a disjoint pair of locksets $(L_+,
L_-)$ (called positive and negative lockset, respectively), where
$L_+$ is the set of \emph{additional} locks that are acquired on
\emph{all} executions from the function entry to the given location,
and the $L_-$ is the set of all locks that may have been released on
\emph{some} execution from the function entry to the given location.
A guarded access corresponds to each \emph{lvalue} in a program, and
it also contains the corresponding relative lockset and the type
(i.e., read or write) of the memory operation. For a function (the
caller function) which calls other functions (the callee functions),
R\textsc{elay} summarizes the relative locksets and the guarded
access sets of all the callee functions to compute the relative
locksets and the guarded access sets of the caller function. Through
traversing the call graph of each thread in a bottom-up manner, the
guarded access set of each memory operation in each thread can be
computed, which can determine potential data race based on whether
or not the corresponding positive lockset is empty.

Figure~\ref{fig:relay} gives an illustrative example about the
bottom-up manner process of R\textsc{elay}. At first, R\textsc{elay}
computes the the relative sets and the guarded access sets of the
functions on the leaves of the call graph, i.e., function
\emph{BlockOwner}, \emph{BlockOwnerColumn} and \emph{BlockOwnerRow}.
Then, R\textsc{elay} summarizes the information of function
\emph{BlockOwner}, \emph{BlockOwnerColumn} and \emph{BlockOwnerRow}
to compute the information of function \emph{lu} and \emph{TouchA}.
Finally, after the computation reaches the thread entry as
\emph{SlaveStart}, R\textsc{elay} intersects the lockset of the
guarded access of each memory operation to find out all potential
races.

\subsection{Race Warnings Reported by Relay}
Although the effectiveness of R\textsc{elay} has been demonstrated
by a Linux kernel 2.6~\cite{Voung07FSE}, here we still evaluate it
on various application benchmarks, to discuss some issues about how
to utilize its race warning report in RacX. The programs employed in
the evaluation are as shown in Table~\ref{tab:benchmark}, including
\emph{apache}, \emph{aget}, \emph{pfscan}, \emph{fft}, \emph{lu},
\emph{water-nsquared}, and \emph{ocean}. These programs cover
different application fields (network, desktop, and scientific
computing) and scales.

\begin{table}[tb]
\caption{Evaluated Parallel
Programs}\label{tab:benchmark}\centering{
\begin{tabular}{|c|c|c|}
\hline \textbf{\emph{Benchmark}} & \textbf{\emph{Size(LoC)}} & \textbf{\emph{Notes}}\\
\hline \hline \emph{apache} & 230K & Network Application\\
\hline \emph{aget} & 2.5K & Network Application\\
\hline \emph{pfscan} &1.0K& Desktop Application\\
\hline \emph{fft} &1.1K& Scientific Computation\\
\hline \emph{lu} &1.2K& Scientific Computation\\
\hline \emph{water-nsqaured} &3.5K& Scientific Computation\\
\hline \emph{ocean} &8.2K& Scientific Computation\\
\hline
\end{tabular}}
\end{table}

For these programs, the number of race warnings reported by
R\textsc{elay} are shown in Table~\ref{tab:warnings}. Apparently, a
race warning always involves many operations from different lines of
code, which results in that the number of \emph{race warning pairs},
i.e., two operations compete for accessing the same memory location,
are much larger than the number of reported race warnings. Besides,
Table~\ref{tab:warnings} also gives the number of \emph{race warning
sites}, i.e., the lines of code where reported races may happen.
From Table~\ref{tab:warnings} we can find that the number of memory
operations involved in the reported potential races of
R\textsc{elay} is already very low compared with the number of all
memory operations in the evaluated program.

However, instrumenting all of potential races reported by
R\textsc{elay} can still be quite inefficient for deterministic
replay. One of the most important reasons is the number of
\emph{race warning instances}, i.e., the occurrences of races in a
concrete execution, can be extremely large when the corresponding
memory location is frequently accessed for many times. It is very
common that some race warnings lie in the hot-spot or critical path
of the program. For example, a false positive on a shared array can
occur millions of times for one execution of program \emph{lu},
which dominates the total number of race warning instances.
Instrumenting it will bring remarkable slowdown to the record run
and the replay run. Therefore, to make RacX more practical, we
should utilize some techniques to reduce the number of false
positives, especially performance-critical ones, as much as possible
without loss of any real data race.

\begin{table}[tb]
\caption{Insight of Race Warnings Reported By R\textsc{elay}
}\label{tab:warnings}\centering{
\begin{tabular}{|c|c|c|c|}
\hline \textbf{\emph{Benchmark}} & \textbf{\emph{Race Warnings}} & \textbf{\emph{Warning Pairs}}&\textbf{\emph{Warning Sites}}\\
\hline \hline \emph{apache} & 4739 & 273258& 11590 \\
\hline \emph{aget} & 46 & 230&108\\
\hline \emph{pfscan} &20 & 93&67 \\
\hline \emph{fft} &26& 281& 145\\
\hline \emph{lu} &16& 104&88 \\
\hline \emph{water-nsquared} &115& 712& 488\\
\hline \emph{ocean} &77& 2678&1807 \\
\hline
\end{tabular}}
\end{table}

\subsection{Analysis of False Positives}
To find out potential solutions to prune false positives for more
precise instrumentation, we analyze the reported potential races of
evaluated programs. We find out that the main sources of false
positives may lie in the following categories:

\begin{itemize}
\item \emph{Unlikely aliasing}: Due to the inherent limitation of static method,
static data race detectors have to conservatively analyze pointer
alias. In some extreme cases, a pointer is even assumed to be able
to access any memory location. It would make the static detector
falsely report race between two non-conflicting memory operations.
Actually, many false positives in \emph{apache} fall into this
category due to the frequently employed function pointers.
\item \emph{Unlocked initialization}: Shared variables can be initialized,
e.g., by function \emph{malloc} or resetting to an initial value,
without obtaining a lock. Thus, intersecting the locksets of a
shared variable during initialization with other accesses to this
variable may cause issuing a potential race. This is an essential
drawback of lockset-based analysis, and this kind of false positives
is pervasive in all evaluated programs according to our analysis.
\item \emph{False sharing on array}: Static data race detectors rely heavily on
point-to analysis (PTA) to find out conflicting memory operations
that access the same variable. However, it is a great challenge for
static analysis to distinguish array elements with different
subscripts. In fact, accessing to different array elements is
typically treated as accessing to the same object as the start
location of this array. For scientific computation programs which
intensively employ array manipulations, such imprecise PTA may lead
to a large number of false positive instances.
\item \emph{Self competence}: To obtain potential races, lockset intersections
are performed on \emph{each} thread entry pairs without considering
whether or not the thread can be spawned more than one time. For
example, program \emph{aget} may spawn only \emph{one} helper thread
\emph{signal\_waiter (sw)} to update the progress bar or handle the
interrupt signal. However, the thread entry pair as $\{sw, sw\}$ is
still statically analyzed assuming that two instances of \emph{sw}
can be executed simultaneously.
\item \emph{Happens-before relation}: Happens-before relation~\cite{Lamport78CACM} indicates that there
exists a partial order between two events (i.e., two accesses to the
same memory location). Once access $A$ happens-before access $B$, no
race occurs between these two accesses. Such a relation is often
causes by \emph{signal/wait}- or barrier-type synchronization, which
cannot be properly handled by the lockset-based analysis. For
example, the main thread often waits for the completion of the child
threads, after that, statistic information about the entire
execution may be printed by accessing shared variables without
holding locks.
\end{itemize}

\begin{figure} [htbg]
\centering
\includegraphics[width=0.5\textwidth]{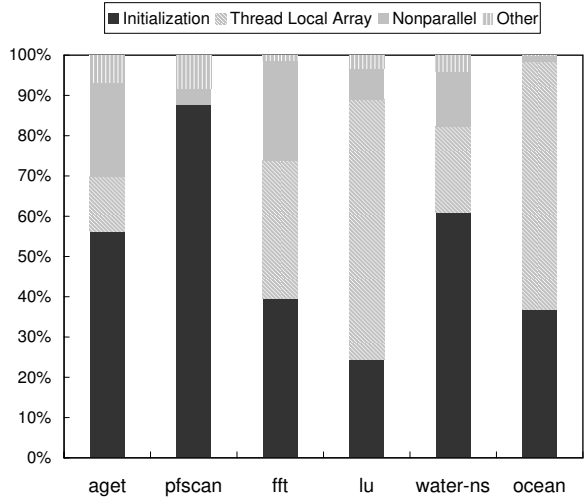}
\caption{Summary of the distribution of different kinds of false
positives according to the number of race warning sites. Since most
false positives of \emph{apache} are unlikely alias, here we only
consider the false positives of left programs, which can be divided
into unlocked initialization, false sharing on array, non-parallel
execution and other (which may contain real data
races).}\label{fig:sum}
\end{figure}

Actually, both \emph{self competence} and \emph{happens-before
relation} can be summarized as \emph{non-parallel execution} of
corresponding (potential) threads. Thus, Figure~\ref{fig:sum}
illustrates the distribution of different kinds of false positives,
more specifically, unlocked initialization, false sharing on array,
non-parallel execution and others, except for \emph{apache} that at
least 90\% false positives are caused by unlikely aliasing. Not
surprising, the distribution of false positive categories varies
significantly among programs. We can clearly see that initialization
is indeed very common for all evaluated programs. For scientific
computation programs, different threads always access different
parts of a global array without holding locks, which results in a
large number of false sharing on array elements. For desktop
application as program \emph{pfscan}, false positives only stem from
unlocked initialization and a small number of non-parallel
execution. For network applications as program \emph{aget}, in
addition to some real data races that fall into the category of
\emph{other}, the number of false positives originated from
non-parallel execution is also considerable.

Considering stated kinds of false positives, \emph{unlikely
aliasing} is the inherent characteristics of static analysis due to
lack of runtime information. Thus, to ensure the completeness of
race report, we cannot prune this kind of false positives. Besides,
for the potential races caused by \emph{self competence}, we cannot
aggressively remove them, since the static detector is often
impossible to know whether or not a thread is spawned for only one
time without precise semantic analysis.

In this work, we especially focus on resolving the false positives
caused by unlocked initialization and false sharing on array by some
automatic techniques.

\textbf{Handling Unlocked Initialization.} Automatically identifying
all initialization codes statically is intractable since it is not
easy to know when initialization is complete. To reduce the number
of false positive caused by unlocked initialization, we first
identify simple yet common initialization pattern by a compiler
pass. Besides, we consider to add locks on initialization code to
avoid add instrumentation codes for every access to this shared
variables. The basic intuition is that initialization always
executes much more infrequently than the subsequence accesses. Thus,
adding locks on the initialization bring trivial overheads compared
with instrumenting all accesses. Section 4 elaborates the detail
mechanism.

\textbf{Handling False Sharing on Array.} It is common that threads
access different parts of a shared array without holding locks by
dedicate partition algorithms in scientific computation. Actually,
for scientific computation programs, the number of race warning
instances caused by false sharing on array is extremely large, which
indicates that executing corresponding instrumented codes dominates
the total execution time. Thus, for these programs, these race
warnings are performance-critical warnings that significantly
impedes the performance of production run and replay run.


Fortunately, the accessed range of array from each thread is always
determined by unique thread ID. Thus, once we can check the
potential range a specific thread touched, and then intersect the
ranges of different threads, we can determine whether or not they
access common array elements. Once the constraint that there is no
common array element accessed by threads with different thread ID is
satisfied, we can eliminate the corresponding false positives.
Section 5 introduces details of our symbolic array cross-range
checking mechanism for achieving this goal.



\section{Initialization Pruning}
\subsection{Initialization Identification}

\begin{figure} [htbg]
\centering
\includegraphics[width=0.5\textwidth]{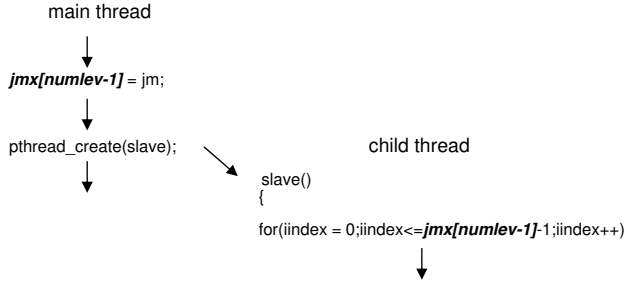}
\caption{A typical initialization pattern extracted from program
\emph{ocean}. The shared array \emph{jmx} is initialized in the main
thread. Then, it is accessed by several child threads forked by the
main thread.}\label{fig:oinit}
\end{figure}

A common initialization pattern of multi-threaded program is as
shown in Figure~\ref{fig:oinit}: After some initialization on shared
variables, the main thread forks several child threads via creation
primitives, e.g., \emph{pthread\_create}, to accomplish a task in
parallel. Based on this observation, we manage to exclude potential
data races between the child threads and the initialization code in
the main thread which are always executed \emph{before} forking
child threads.

We utilize a compiler pass to automatically identify this
initialization pattern in the source code. First, the function $F$
that calls thread creation functions are identified. Then, by
iterating all statements of function $F$, we can find the
\emph{dominators}\footnote{We say that a statement D is a dominator
of another statement A if every execution path from the start to A
must go through D.} of the statement calling the thread creation
function which spawns a child thread (denoted as $T$). Apparently,
memory operations in these dominators will not contend with those in
thread $T$. Therefore, corresponding race warnings must be false
positives. Although this initialization identification criterion is
simple and conservative, it can avoid to instrument some
``hot-spot'' operations, thus can effectively reduce the
instrumentation overhead for some scientific computation programs as
demonstrated by our experiments in Section~\ref{sec:experiments}.

\subsection{Locking Possible Initialization}
Other than the pattern mentioned in the previous subsection, there
are also many other ad hoc initialization patterns in practice. It
is quite hard to statically identify \emph{all} initialization code
from the source codes. However, we observe that a potential race
about initialization often involves \emph{only one unlocked
operation about a shared variable and N other locked operations
about the shared variable}. Hence, for each potential race with only
one unlocked operation, we add a lock on the unlocked operation to
resolve the race warning, which can avoid to impose instrumentation
codes to all of the $N+1$ operations involved in the race warning.

In fact, we cannot ensure that the only unlocked operation involved
in a race warning is a real initialization. In other words, even
real data races may also have this property (involving only one
unlocked operation). That is the reason for why we call such
unlocked operation as \emph{possible} initialization. Fortunately,
even for a real data race, adding lock on this possible
initialization can still guarantee successful replay, since all
operations involved in the race have been well guarded by
synchronization operations, whose orders are also recorded by RacX.

\section{Array Cross-Range Check}
Due to the limitation of existing PTA and conservative property of
static analysis, array objects are always treated as an entirety
regardless of its concrete subscripts. Thus, for scientific
computation programs with intensively accessing to shared arrays,
many false positives would be reported by static detector. In this
section, we introduce the \emph{symbolic array cross-range checking}
technique to handle this kind of false positives.

\subsection{Basic Idea}
In scientific computation, it is common that several threads access
a shared array without holding locks since the algorithm guarantees
each thread to access a specific array region without overlapping.
Based on this observation, for a reported race warning about a
shared array, we manage to analyze which range of array is accessed
by each thread, and check whether there is a common range of array
accessed by multi threads. The concrete flow of symbolic array
cross-range checking includes three steps as follows:

In the first step, we conduct the \emph{intraprocedural range
analysis} on each function in a program, which produces the
\emph{local range constraint} for each program statement in the
function. For example, the range constraint for a statement such as
$for(i = 0; i < 10; i++)$ is $0 \le i \le 9$. The intraprocedural
range analysis can be implemented with Purdue's compiler analysis
tool as Cetus~\cite{Dave09Computer}, which is a source-to-source
compiler infrastructure for C programs.

The local range constraints produced by the intraprocedural range
analysis are a series of expressions of local variables. To
facilitate inter-thread analysis, the second step needs to produce
the \emph{global range constraint}, which associates the local
variables in the range constraints with the expressions of the
global variables and the variables in the thread entry function.
Hence, we develop an in-home tool to implement the
\emph{interprocedural range analysis}, which can traverse the call
graph (CG) from the entry function of each thread, and iteratively
associate the local variables in the range constraints of callee
function with the variables in the caller function and the global
variables. After the whole traversal, the tool can output the set of
global range constraints for all array accesses in each thread.

Based on the result of interprocedural range analysis, we can carry
out the third step, which checks whether there is some element of a
shared array accessed by multi threads with a given statement. We
treat the checking problem as a constraint satisfaction problem
(CSP) problem, and employ a complete CSP solver
STP~\cite{Ganesh07CAV} to solve the CSP problem. A typical CSP
problem consists of an objective expression and a set of constraints
on the variables in the objective expression. In our checking
problem, the objective expression aims at determining whether or not
the array subscript expression in the statement with different
thread identifiers can be satisfied to equal. Meanwhile, the set of
constraints on variables in the objective expression can be obtained
directly from the set of global range constraints produced by the
interprocedural range analysis. If the objective expression is
proven to be unsatisfiable under the constraints, there is no array
element accessed by multi threads with the statement, which can
eliminate the possibility that the statement has a data race with
itself about the array.

\subsection{An Illustrative Example}
\begin{figure}[htbg]
\centering
\includegraphics[width=0.6\textwidth]{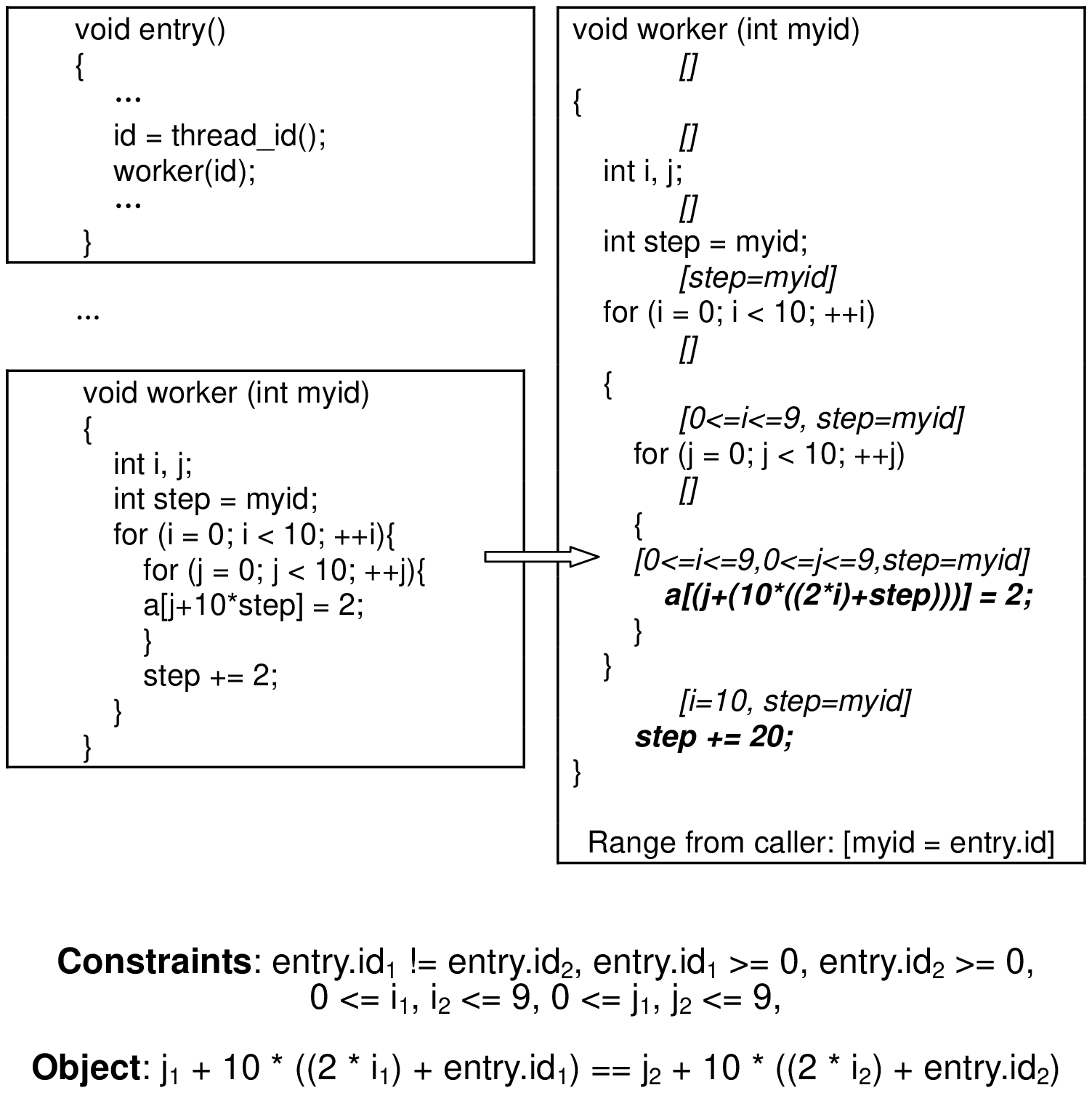}
\caption{An illustrative example to demonstrate the symbolic array
cross-range check for find-grain inspection of racing array access
between threads.}\label{fig:example}
\end{figure}

An illustrative example on how to conduct array cross-range checking
is as shown in Figure~\ref{fig:example}. In a typical multi-threaded
program, function \emph{entry} is created as the entry point
function for each thread. In each thread, function \emph{entry}
stores its own thread id in variable $id$ and calls function
\emph{worker} to accomplish computation on the shared array $a$. Our
aim is to determine whether some element of $a$ is accessed by multi
threads.

As the first step, we perform the intraprocedural range analysis on
function \emph{worker} to obtain the local range constraints for
each statement. These local range constraints are written in
\emph{[]} in the right side of Figure~\ref{fig:example}. For
example, for accesses to array \emph{a} in the loop, we can find its
local range constraints $R$ as $\{0 \le i \le 9,0 \le j \le
9,step=myid\}$.

Although the local range constraints of function \emph{worker}
restrict the possible values of the variables in the subscript
expression of $a$ (such as $i$, $j$, and $step$), we are not aware
of that the value of $step$ is depend on the thread identifier $id$
which appears in the caller function \emph{entry}. Thus, we employ
the second step as interprocedural range analysis to obtain the
global range constraints as $\{0 \le i \le 9,0 \le j \le 9,
step=entry.id\}$ by associating the local variables $step$ and
$myid$ with the variable $entry.id$ from  the caller function
$entry$ (here we use the name of caller function as the prefix to
indicate that the variable is passed from function $entry$).

After getting the global range constraints, we can formulate the
array cross-range checking as a CSP to determine whether or not
different threads can access common elements. In this example, the
object expression is $j_1+10*(2*i_1+entry.id_1)==
j_2+10*(2*i_2+entry.id_2)$, while the global range constraints are
\{$0 \le i_1,i_2 \le 9,0 \le j_1,j_2 \le 9, entry.id_1 \ge
0,entry.id_2 \ge 0, entry.id_1\neq entry.id_2$\}. By querying the
STP, we confirm that the object expression is unsatisfiable under
the global range constraints. Hence, there does not exist a pair of
threads which access the same element of array $a$, which indicates
that race warnings about array $a$ can be excluded.

It is worth noting that before the range analysis process, we need
to perform the so-called induction variable substitution (IVS) on
the original program. In this example, variable $step$ is not
directly controlled by explicit variable as $i$ and $j$, which
brings additional difficulties to the range analysis. If we directly
process this function, we might obtain the set of range constraints
such as $\{0 \le i \le 9,0 \le j \le 9, step \ge entry.id\}$. By
solving the CSP transformed from subscript expression as $j + 10 *
step$, we cannot ensure that different threads access different
parts of this array. However, after performing IVS on the original
program, induction variable \emph{step} becomes a fix value (i.e.,
myid) during the entire execution of this loops, and the subscript
expression can be rewritten as $j+10*(2*i+step)$, which greatly
facilitates range analysis.

\section{Record/Replay Implementation}
As mentioned, to deterministically replay a multi-threaded program,
both synchronization and data race orders are recorded in RacX. In
this section, we introduce the record/replay mechanism for these two
kinds of orders in RacX.

\subsection{Recorder}
During the recording phase, we need to record the execution orders
among synchronization and data races. For each synchronization
(e.g., \emph{pthread\_mutex\_lock}, conditional wait, barrier, and
so on), we record its timestamps of Lamport
clock~\cite{Lamport78CACM} at the pthread library level similar with
several other deterministic record/replay schemes such as
RecPlay~\cite{Ronsse99TCS} and PRES~\cite{Park09SOSP}. Although our
scheme may miss some races in low level library, e.g., \emph{glibc},
in practice, recording synchronization at the pthread library level
is sufficient to successfully replay a multi-threaded program, while
its performance overhead is much smaller than of recording at the
glibc library level. For the notoriously intractable data races, we
also utilize Lamport clock to trace race orders. In addition to
recording and increasing the timestamp for each shared memory
access, we also have to log the instruction count of traced memory
accesses for facilitating replay.

Moreover, to simplify the recording process, here we only utilize
two global clocks to log the timestamps of synchronization
operations and data races, respectively. Although we could set one
global clock for each synchronization operation related to the same
synchronization variable (or for each data race accessing the same
memory location) to allow more parallelism in replay phase, we found
that the replay performance cannot be improved significantly since
the replay overheads are already quite small.

\subsection{Replayer}

During the replay phase, two global variables are employed to track
current execution clock. We also maintain a single global variable
to track current instruction count for each thread, which servers as
an index to query the timestamp of current synchronization operation
or data race from the recording trace. Thus, in order to reproduce
race orders, e.g., data race orders, once encountering a memory
access that may cause a potential race, replayer should query the
loaded trace from the recording phase to obtain the timestamp of
this access based on current instruction count. Then, if
corresponding clock equals to its timestamp, thread $T$ can proceed.
Otherwise, $T$ should wait for its turn.


\section{Evaluation}~\label{sec:experiments}
\subsection{Experiment Setup}
We conduct experiments on a 2.13GHz 16-core Intel Xeon SMP system
with 8GB RAM and Linux version 2.6.18. The programs we employed for
evaluation are listed in Table~\ref{tab:benchmark}, including
network application, desktop application and scientific computation
from SPLASH-2~\cite{Woo95ISCA}. For each program, we measure the
performance results of the original execution (without any source
code modification), the record run, and the replay run. Moreover,
each performance result is the average performance of \emph{five}
trials (which follow an excluded warm-up trial) to resist the common
speed fluctuation of multi-threaded programs.

To evaluate the performance results of program \emph{apache}, we
utilize ab (Apach Bench) to send 500 requests from 16 concurrent
client to the server over a local network. We evaluate program
\emph{aget} by retrieving a 53MB file from a server over a local
network. For program \emph{pfscan}, we use it to search a string in
a directory with 968MB log files in parallel. 
For program \emph{fft}, we set the parameter $m$ as 22, where there
are $2^m$ complex data points transformed. And we set the parameter
$n$ for program \emph{lu} and program \emph{ocean} as 1600 and 1026,
respectively. For program \emph{water-ns}, the number of molecules
is 2197. To evaluate the scalability of RacX, for each program, we
vary the number of worker threads as 1, 2, 4, 8 and 16 (however, for
\emph{aget}, the maximal number of threads cannot exceed 10, so we
confine the maximal number of its worker threads to 8). Besides,
here the number of worker threads are threads for effective
computation, e.g., \emph{get} threads in \emph{pfscan} and download
threads in \emph{aget}.

\subsection{Recording Overheads of Instrumenting All Race Warnings}
In contrast to previous work that instrumenting all shared locations
for recording the global orders of shared access to the same
location from different
threads~\cite{Narayanasamy06SIGMETRICS,Patil10CGO}, the number of
instrumented memory locations might be greatly reduced since only
shared memory accesses that are relevant to potential races would be
instrumented. Thus, in this experiment, we instrument all race
warning sites in the raw report of a state-of-the-art race detector,
which is named R\textsc{elay}, and compare its performance with the
original execution. Obviously, if the recording overheads are very
low, we can directly employ the raw race report of static race
detectors for instrumentation without any post-process. Otherwise,
false positive pruning techniques should be utilized to reduce
instrumentation efforts.

\begin{figure}[htbg]
\centering
\includegraphics[width=0.5\textwidth]{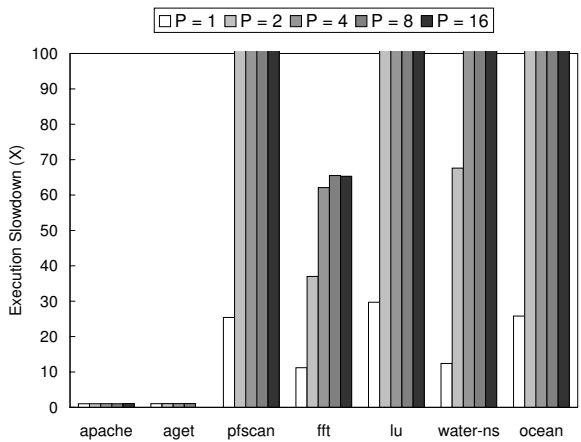}
\caption{Execution slowdown of the recording run to the original run
when instrumenting all race warnings reported by R\textsc{elay}. The
recording slowdowns vary significantly among programs. Two extreme
cases are \emph{apache} and \emph{lu}, whose recording slowdowns are
1.08X and 900X respectively when running with 16
threads.}\label{fig:raw}
\end{figure}

Figure~\ref{fig:raw} shows the execution slowdown of the recording
run to the original run when instrumenting all reported race
warnings. It is obvious that among the evaluated programs, the
recording overheads for program \emph{apache} and program
\emph{aget} are already very low (only 1.08X and 1.07X respectively)
even when running with 16 threads. Taking program \emph{apache} as
an example, the reason may be that many instrumented codes are
seldom executed, e.g., configurations of modules. Thus, program
\emph{apache} server only executes a small part of all instrumented
codes during processing the requests from clients (generated by
apache bench).

On contrary, for some desktop programs and scientific computation
programs, the recording overheads are too large to switch on
recording all race warnings in the raw report. For instance, when
running with 16 threads, the maximal execution slowdown could even
exceed 900X for program \emph{lu} (exceeding the upper bound of
Figure~\ref{fig:raw}). Even for program \emph{fft} with the least
slowdown as 11.2X when running with 2 threads, it is still far from
ideal deterministic replay approaches. In fact, for these programs,
instrumentation are apparently implanted to the false positives in
the hot-spots or critical paths, which have critical impacts on the
performance. Therefore, pruning the false positives in the raw race
report is very crucial to the successful of RacX.

\subsection{Effects of False Positive Pruning}
To reduce the instrumentation overheads, our \emph{Analyzer} employs
the proposed false positive pruning techniques, i.e., initialization
pruning and array cross-range checking as introduced in previous
sections, to exclude the ``critical'' false positives reported by
R\textsc{elay}.

\begin{figure}[htbg]
\centering
\includegraphics[width=0.5\textwidth]{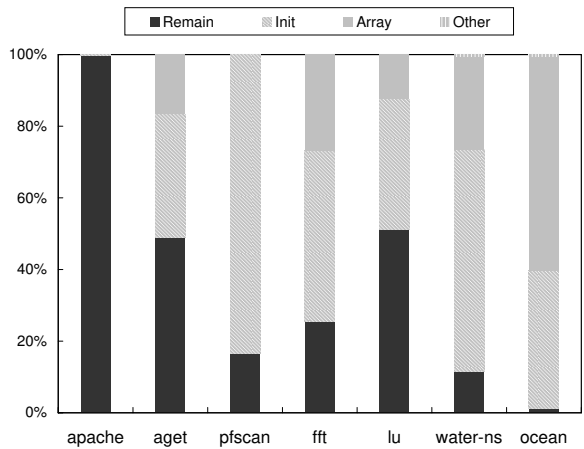}
\caption{Effects of proposed false positive pruning techniques on
evaluated programs with respect to \emph{race sites}. Expect for
program \emph{apache}, our techniques can reduce race warning sties
for 49\% (\emph{lu}) to 98\% (\emph{ocean}), which significantly
reduces the memory operations to be instrumented.}\label{fig:filter}
\end{figure}

According to Figure~\ref{fig:filter}, we can see that initialization
pruning is effective to reduce race warnings for all evaluated
programs except for program \emph{apache}. It can reduce race
warning sites for 34\% (program \emph{aget}) to 84\% (program
\emph{pfscan}). Meanwhile, for all scientific programs, array
cross-range checking is of great significance, as evidenced by that
it can reduce race warning sites for 13\% (program \emph{lu}) to
60\% (program \emph{ocean}). However, it falls short of reducing
false positives of other programs such as program \emph{pfscan}
since this kind of program does not have any shared array
computation involved in reported race warnings.


To sum up, out techniques significantly reduce the number of race
warnings in the refined report for most programs (from $49\%$ for
\emph{lu}) to $98\%$ for \emph{ocean}). The only exception is
program \emph{apache}\footnote{The ineffectiveness of our techniques
on \emph{apache} is caused by that the most false positives of
\emph{apache} are unlikely alias which are hard to analyze by static
methods}, which has already negligible performance overhead even
instrumenting all race warnings in the raw report. As a result, the
race warning sites in the refined report become an extremely small
portion of all the memory operations in multi-threaded program. As
shown by \emph{Refined Sites} column in Table~\ref{tab:comparison},
most programs contain less than $100$ race warnings. Program
\emph{ocean} even has only $24$ race warnings. It is not hard to
imagine the low efforts to instrument these few race warnings in the
\emph{Recorder} and \emph{Replayer} of RacX.

\begin{table}[tb]
\caption{Reduced race warning sites for instrumentation in
RacX}\label{tab:comparison}\centering{
\begin{tabular}{|c|c|c|}
\hline \textbf{\emph{Benchmark}} & \textbf{\emph{Raw Report}} & \textbf{\emph{Refined Report}}\\
\hline \hline \emph{apache} &  11590 &11546\\
\hline \emph{aget} & 108&53\\
\hline \emph{pfscan} &67&3\\
\hline \emph{fft} &145&37\\
\hline \emph{lu} &88 &45\\
\hline \emph{water} &464& 53\\
\hline \emph{ocean} &1807& 24\\
\hline
\end{tabular}}
\end{table}

\subsection{Record/Replay Performance of RacX}
The overall performance results of RacX are shown in
Table~\ref{tab:racx}. The fist three columns show the name of
evaluated programs, the number of working threads and the original
execution time (average of $5$ trials). The next two columns give
the execution time when running in recording mode (average of $5$
trials) and corresponding overheads to original run. Similarly, the
last two columns show the execution time (average of $5$ trials) and
corresponding overheads when running in replay mode.

\begin{table}
\caption{Performance of RacX when running with record- and replay-
mode compared with original execution. We vary the number of threads
from 1 to 16 to demonstrates the scalability of
RacX.}\label{tab:racx}\centering{
\begin{tabular}{|c|c|c|c|c|}
\hline Benchmark & Worker & Original  &  Recording & Recording \\
 &  Threads &Time (s) &  Time(s) &  Overheads \\
\hline \hline \emph{apache}&1&18.36&18.51&0.81\%\\
 \cline{2-5} &2&17.89&18.09&1.12\%\\
 \cline{2-5} &4&16.18&16.82&3.90\%\\
 \cline{2-5} &8&14.38&14.73&2.48\%\\
 \cline{2-5}  &16&11.32&12.07&6.65\%\\
\hline \hline \emph{aget}&1&9.06&9.46&4.43\%\\
\cline{2-5}  &2&5.82&6.20&6.40\%\\
\cline{2-5}  &4&5.83&6.24&6.98\%\\
\cline{2-5}  &8&5.87&6.33&7.76\%\\
\hline \hline \emph{pfscan}&1&5.38&5.38&0.12\%\\
\cline{2-5}  &2&2.80&2.81&0.14\%\\
\cline{2-5}  &4&1.44&1.45&0.15\%\\
\cline{2-5}  &8&0.78&0.78&0.28\%\\
\cline{2-5}  &16&0.49&0.49&0.57\%\\
\hline \hline \emph{fft}&1&3.67&3.69&0.56\%\\
\cline{2-5}  &2&2.24&2.25&0.73\%\\
\cline{2-5}  &4&1.51&1.54&1.67\%\\
\cline{2-5}  &8&1.18&1.19&0.98\%\\
\cline{2-5}  &16&1.04&1.05&1.58\%\\
\hline \hline \emph{lu}&1&9.66&9.69&0.25\%\\
\cline{2-5}  &2&5.16&5.08&-1.53\%\\
\cline{2-5}  &4&2.79&2.73&-1.97\%\\
\cline{2-5}  &8&1.63&1.59&-1.90\%\\
\cline{2-5}  &16&1.05&1.04&-0.99\%\\
\hline \hline \emph{water-ns}&1&6.21&6.26&0.74\%\\
\cline{2-5}  &2&3.37&3.4&1.03\%\\
\cline{2-5}  &4&1.91&1.94&1.66\%\\
\cline{2-5}  &8&1.23&1.28&4.15\%\\
\cline{2-5}  &16&1.11&1.54&38.7\%\\
\hline \hline \emph{ocean}&1&6.66&6.68&0.22\%\\
\cline{2-5}  &2&3.30&3.31&0.56\%\\
\cline{2-5}  &4&1.68&1.70&0.87\%\\
\cline{2-5}  &8&1.11&1.13&1.62\%\\
\cline{2-5}  &16&0.91&0.95&3.92\%\\
\hline
\end{tabular}}
\end{table}

In recording mode, the average overheads across all evaluated
programs are 1.21\%, 1.89\%, 2.20\%, and 8.41\% when running with 2,
4, 8 and 16 threads (except for program \emph{aget} that cannot be
executed with 16 threads), respectively. Thus, we can conclude that
RacX is efficient and scalable in practical usages (e.g., debugging,
fault tolerance, intrusion detection, and so on), since RacX works
well on large scale commercial program and 16 worker threads. Among
these programs, the most surprising one might be \emph{lu} since the
recording run is even slightly \emph{faster} than the original run.
According to Table~\ref{tab:comparison}, there are still $45$ race
warning sites are still left for instrumentation, however, most race
warning sites have only one instance (i.e., memory operations
related to these race warning sites are executed for only one time).
Therefore, these instrumentation codes have little impacts on the
original execution. Furthermore, by comparing the execution details
of the recording run and the original run, we find that the barrier
time of the recording run is much smaller than the original run
since the execution of worker threads are more balanced due to the
import of instrumentation, which reduces the overall waiting time.

\section{Related Work}~\label{sec:related}

\noindent\textbf{Deterministic Record/Replay.} Deterministic
record/replay has been extensively investigated by various research
communities, including computer architecture, operating system,
parallel programming and debugging, reliability, hardware
verification, and so on.

Initially, deterministic record/replay schemes focus on providing
solutions for replica execution on uni-processor systems. These
schemes are implemented in different abstract levels to record the
non-deterministic events on uni-processor systems (e.g., I/O inputs,
interrupts), e.g., Jockey~\cite{Saito05AADEBUG},
Liblog~\cite{Geels06ATC} and R2~\cite{Guo08OSDI} in library level,
ReVirt~\cite{Dunlap02OSDI}, TTVM~\cite{King05ATEC} and
Nirvana~\cite{Bhansali06VEE} in virtual machine level, and
Flashback~\cite{Srinivasan04ATC} in system call level.

Deterministic record/replay schemes for multi-processor systems need
to confront an additional type of non-deterministic events: data
races. Informally, data race exists between a pair of conflicting
memory operations without synchronization. It may make corresponding
memory operations produce different results in different executions.
Although the amount of data races is not high in real-world
programs, it is extremely difficult to identify them: There are
enormous number of pairs of conflicting memory operations, while
identifying the existence of data race between even a pair of memory
operations is also NP-hard~\cite{Netzer90ICPP}. Hence, previous
investigations on deterministic record/replay do not attempt to find
out all the data races and record them. On the contrary, they adopt
two types of substitution strategies: Recording \emph{all} memory
operations, and recording \emph{none} memory operation.

The intuition between the first strategy is that once the orders of
all memory operations are recorded, the orders of all data races are
also recorded. In addition to InstantReplay~\cite{LeBlanc87TC} that
records the orders of shared objects at a coarser level.
SMP-ReVirt~\cite{Dunlap08VEE} is a representative scheme the first
strategy that can deal with fine-granularity shared memory access.
It relies on the OS page management mechanism to record the ordering
of all conflicting memory operations. It may lead to $1.5X$-$8.7X$
slowdown to the production run. Another scheme
PinPlay~\cite{Patil10CGO} employs a dynamic instrumentation tool
(Pin~\cite{Luk05PLDI}) to monitor all memory operations. It incurs
$36X$-$146X$ slowdown to the production run.
DoublePlay~\cite{Veeraraghavan11ASPLOS} can convert the orders
between memory operations to the orders of thread scheduling through
speculative execution, thus it only needs to record the thread
scheduling. However, it has to consume additional cores (the number
of additional cores must equal to the number of cores in regular
execution) to provide speculative results for reference, which may
reduce the overall throughput of the system for $50\%$. Since
recording all memory operations with only software brings remarkable
performance overheads, some hardware-assisted schemes were proposed
to reduce the performance overhead (as well as log size) of
recording with non-standard hardware
supports~\cite{Xu03ISCA,Hower08ISCA,Montesinos08ISCA,Voskuilen10ISCA}.
However, none of these non-standard hardware supports has been
adopted by industry yet, hence these hardware-assisted schemes
cannot work on commodity systems.


The second strategy does not record any memory operation to avoid
the performance overhead of the first strategy. Instead, it only
records the synchronization orders and a few other information
(program outputs, execution path, and so on).
RecPlay~\cite{Ronsse99TCS} is the first scheme adopting the second
strategy. It has no knowledge about the normal memory operations,
thus fails to replay the programs with data races.
PRES~\cite{Park09SOSP} does not record all memory operations. To
provide determinism, it has to repeatedly replay the program, until
that some replay run happens to output the same with the production
run. However, there is not theoretical upper bound for the replay
times to meet the first satisfactory replay run.
ODR~\cite{Altekar09SOSP} proposes to infer all the memory operation
orders from the recorded program outputs, synchronization orders,
and execution path information. However, it may fall short of some
corner-cases due to the state exploration problem of inference.
Respec~\cite{Lee10ASPLOS} employs speculative execution to alleviate
the burden of deterministic replay by only recording synchronization
operations. However, it does not support off-line replay, thus
cannot be employed by debugging multi-threaded programs. Besides, it
needs additional cores to provide speculative results for reference
similar with DoublePlay~\cite{Veeraraghavan11ASPLOS}, which may
reduce the overall throughput of the system for $50\%$.

Furthermore, all schemes adopting the second strategy must relax
their fidelity due to the lack of the order information of memory
operations. They can only provide the so-called output determinism,
which is a relaxation of the value determinism provided by RacX. The
output determinism offers no guarantee on the identity of the
execution details among the production run and the replay run. Such
low level of determinism cannot fulfill the requirements of many
important applications of deterministic record/replay (such as
intrusion detection and workload capture), which care not only the
outputs but also the execution details.

In general, RacX is substantially different with the previous
deterministic record/replay schemes, since it is the first scheme
that directly records and replays data races. From the perspective
of soundness, RacX can faithfully record/replay multi-threaded
programs with the value determinism, since it monitors all real data
races and synchronization. From the perspective of efficiency, RacX
exhibits the smallest slowdown among all deterministic record/replay
schemes for multi-processor systems as far as we known. From the the
perspective of scalability, RacX exhibits excellent scalability with
respect to the program size and the number of threads. From the the
perspective of portability, RacX has no requirements on the library,
operating system, and the hardware, thus can be easily ported to
various commodity systems.

\noindent\textbf{Deterministic Execution.} Recently, researchers
have proposed deterministic execution as an alternative approach to
achieve determinism without logging
~\cite{Olszewski09ASPLOS,Bocchino09OOPSLA,Devietti09ASPLOS,Bergan10ASPLOS,Devietti11ASPLOS,Hower11HPCA}.
However, the existing deterministic execution methods must impose
predefined orders on the memory operation interleavings in the
execution of multi-threaded program. Hence, these methods incur
relatively high slowdowns to the production run, especially on
commodity multi-processor systems which do not provide non-standard
hardware support for deterministic execution (e.g., 1.1X-6X in
CoreDet~\cite{Bergan10ASPLOS}).

\noindent\textbf{Data Race Detection.} Data race detection can be
categorized into dynamic detectors and static detectors.

A typical dynamic data race detector instruments a program at
runtime and detects data race appeared in the execution of the
program~\cite{Erickson10OSDI}. The concrete detection algorithm can
be based on \emph{happens-before
analysis}~\cite{Lamport78CACM,Choi91TPOLS} (which can be further
improved with respect to efficiency and accuracy by hardware-based
approaches as~\cite{Prvulovic06HPCA,Muzahid09ISCA,Lucia10ISCA}),
\emph{lockset algorithm}~\cite{Savage97TCS}, or
both~\cite{Callahan03PPoPP}. Although dynamic detectors are fast and
ease to use in practice (i.e., reporting a small number of false
positives), they cannot prove the absence of races since their
scopes are limited to one execution with specific inputs and
execution schedule. Even the program has been executed for a large
number of times, we still cannot conclude that all real data races
have been identified. Thus, the completeness requirement of RacX
forces us seeking help for static analysis.

Typically, static data race detectors hunt potential races by
analyzing the source code instead of directly executing the program.
They can use type-based
analysis~\cite{Flanagan00PLDI,Boyapati02OOSPLA,Pratikakis06PLDI} and
data-flow analysis~\cite{Engler03SOSP,Naik06PLDI,Voung07FSE} to
reason the relationship between memory operations in any possible
execution, thus most static detectors are able to find all potential
data races. In contrast to dynamic detectors, the main limitation of
static detectors is that they may produce a large number of false
positives due to the inherent lack of runtime information, such as
pointer alias, array index information, and so on. An user of static
detector often needs to identify the real data races from thousands
of (even more) false positives, which is a tedious job. Although
RacX does not need to know which concrete potential race is real
race, which is false positive, false positives may reduce the
efficiency of RacX. Therefore, RacX employs several automatic
techniques (i.e., initialization pruning and array cross-range
checking), to reduce the number of ``critical" false positives that
lie in hot-spots or critical paths.
%

\section{Conclusions}~\label{sec:conclusion}
In this paper, we propose an efficient deterministic replay scheme,
i.e., RacX, that leverages race-guided instrumentation to
significantly reduce the instrumented memory operations. Since most
shared memory operations are well guarded by synchronization
operations, instrumenting the potential data races (including all
real data races and a few false positives) only brings little
performance overheads to the production run and the replay run.
Furthermore, we propose several automatic techniques, i.e.,
initialization locking and symbolic array cross-range checking, to
prune the false positives which may significantly affect the
efficiency of RacX. Experimental results on several programs with
different scales demonstrate the soundness, effectiveness,
efficiency and scalability of RacX. We hope that these technical
merits of RacX can push the industrial acceptance of deterministic
record/replay for multi-processor systems.

\small{

}
\end{document}